\documentclass[tightenlines,floats,aps,amsmath,amssymb,
prd,showpacs,twocolumn]{revtex4}

\usepackage{amsmath,amssymb,amsfonts}
\usepackage{graphicx}
\usepackage{psfrag}
\usepackage{enumerate} 
\usepackage{calc}

\usepackage{hyperref}

\newcommand{\rd}{{\rm d}}
\newcommand{\Hilk}{{\mathcal{H}}^{\rm kin}}
\newcommand{\Hilp}{{\mathcal{H}}^{\rm phy}}
\newcommand{\re}{\mathbb{R}}
\newcommand{\lPl}{l_{\rm Pl}}

\begin{document}

\title{Big Bounce and inhomogeneities}
\author{David Brizuela}
\email{brizuela@iem.cfmac.csic.es}
\author{Guillermo A. Mena Marug\'an}
\email{mena@iem.cfmac.csic.es}
\author{Tomasz Pawlowski}
\email{tomasz@iem.cfmac.csic.es}
\affiliation{Instituto de Estructura de la Materia,
CSIC,
  Serrano 121-123, 28006 Madrid, Spain
}

\begin{abstract}
The dynamics of an inhomogeneous universe is studied
with the methods of Loop Quantum Cosmology as an
example of the quantization of vacuum cosmological
spacetimes containing gravitational waves
(\emph{Gowdy} spacetimes). The analysis performed at
the effective level shows that: $(i)$ The initial Big
Bang singularity is replaced (as in the case of
homogeneous cosmological models) by a Big Bounce,
joining deterministically two large universes, $(ii)$
the universe size at the bounce is at least of the
same order of magnitude as that of the background
homogeneous universe, $(iii)$ for each gravitational
wave mode, the difference in amplitude at very early
and very late times has a vanishing statistical
average when the bounce dynamics is strongly dominated
by the inhomogeneities, whereas this average is
positive when the dynamics is in a near-vacuum regime,
so that statistically the inhomogeneities are
amplified.
\end{abstract}

 \maketitle

The cosmological models based on classical General
Relativity (GR) predict that the Big Bang singularity
is the true beginning of the universe (boundary of the
spacetime). This prediction, however, is believed not
to be physical since, when one describes the early
universe, GR is applied beyond its domain of validity.
The quantum effects which dominate in this epoch are
expected to resolve the singularity.

This issue was studied recently in the context of Loop
Quantum Cosmology (LQC) \cite{abh} (see also
\cite{kev}). The analysis of a simple model of a
homogeneous and isotropic universe revealed surprising
results \cite{aps}: a large classical expanding
universe was preceded by a(n also large and classical)
contracting one, which bounced (deterministically) and
started to expand once the energy density of its
matter content reached the Planck scale. The
robustness of this result in more general situations
was confirmed later \cite{gen,chiou}, including the
case of anisotropic universes.

However, up to now these studies were mostly
restricted to homogeneous spacetimes. Although a
preliminary analysis of the effects of inhomogeneities
was presented in \cite{boj-inh}, it was not known
whether the modifications to the universe dynamics
predicted by LQC survive in the presence of
inhomogeneities. The possibility to answer this
question arose when an LQC quantization scheme was
formulated \cite{gmm} for a class of cosmological
spacetimes known as Gowdy universes \cite{gowdy}.
These spacetimes, while still symmetric (they admit
two spatial symmetries) include inhomogeneities that
can be interpreted as (linearly polarized)
gravitational waves, thus having local degrees of
freedom. To describe them, a {\it hybrid} quantization
scheme was applied: first the geometry was represented
as the Fourier modes of a gravitational field (the
linearly polarized wave) propagating in a homogeneous
(Bianchi I) spacetime, next this Bianchi geometry was
quantized using loop techniques, while for the
gravitational wave modes standard Fock quantization
methods were employed after introducing a suitable
time-dependent scaling.

This construction paved the way to analyze the quantum
dynamics of this inhomogeneous system; nonetheless the
(field-like) complexity of the quantum configuration
space makes the investigation of the genuine quantum
evolution extremely difficult. This forces one to
resort to {\it classical effective dynamics}
\cite{eff,taveras} -- a classical theory which
incorporates the main effects of spacetime
discreteness and which has been shown to accurately
mimic the quantum evolution in the cases where it has
been tested on so far. Here, we apply this technique
to analyze the quantum dynamical behavior of the Gowdy
universe, answering in particular the following
questions: $(i)$ Does the Big Bounce persist in the
presence of inhomogeneities? $(ii)$ If the answer is
in the affirmative, does the Big Bounce occur in
similar conditions as in homogeneous models? $(iii)$
And how does the structure of the gravitational wave
modes change through the bounce?

Let us start by specifying in more detail the physical
system which we consider here. The Gowdy universes are
vacuum spacetimes with compact sections of constant
time which, in spite of possessing considerable
symmetry, still contain local degrees of freedom.
Namely, these spacetimes possess two spatial
isometries (two commuting spacelike Killing vectors).
The Gowdy universes can be classified by their spatial
topology \cite{gowdy}. The best studied case, on which
we will concentrate, is that with the topology of a
three-torus. This family of spacetimes provides a
generalization of the homogeneous and anisotropic
Kasner solution (for spatially flat topology) to
include inhomogeneities which depend only on one
spatial coordinate \cite{berger}. The spacetime
inhomogeneities can be interpreted as gravitational
waves propagating in an homogeneous background
spacetime. We will consider exclusively the simplest
case of linearly polarized waves, in which, after a
suitable gauge fixing, all the inhomogeneities can be
described by a single metric field and the metric
adopts a diagonal form globally \cite{gmm,gauge}. That
field can be expanded in Fourier series exploiting the
periodicity in the only spatial coordinate on which it
depends (this coordinate is cyclic for the studied
topology). Strictly speaking, the inhomogeneities are
determined by the nonzero modes of this decomposition.
The rest of gravitational degrees of freedom of the
system describe a homogeneous universe on which the
gravitational waves propagate. Specifically, this is a
Bianchi I spacetime with three-torus topology. The
classical solutions of this cosmological model are
known in exact form, and generically present a Big
Bang singularity \cite{mon}.

In order to quantize the system we follow the
prescription used in \cite{aps}, where the geometry
(homogeneous) degrees of freedom were quantized via
loop techniques, whereas for the ``matter'' ones (in
this case the gravitational waves with a convenient
field parametrization) standard (Schroedinger-like)
methods were employed \cite{gauge,unigow}. The
classical metric of the Bianchi I background in the
adopted gauge is diagonal and determined by a triple
of scale factors $a_i(t)$, where $i=1,2,3$. In the
formalism used for quantization, the phase space for
this background is coordinatized by the $SU(2)$
(Ashtekar) connection $A_a^i=c^i\delta^a_i/(2\pi)$ and
the densitized triads $E^i_a=p_i\delta^i_a/(4\pi^2)$
\cite{mmp}. All the information about the system is
encoded in the canonically conjugate variables
$(c^j,p_i)$, where $|p_i|=
|\epsilon^{ijk}|a_ja_k/(8\pi)$ ($\epsilon^{ijk}$ is
the completely antisymmetric unit tensor). Among the
constraints that GR imposes on the system, only two
are not automatically satisfied in the introduced
gauge: the generator of $S^1$ translations on the
inhomogeneous spatial direction (which affects
exclusively the inhomogeneities described by the
gravitational waves) and the spatial average of the
Hamiltonian constraint. The latter can be written (up
to a global constant) as $C=C_{\rm
BI}+C_{\mathcal{F}}$ (where $C_{\rm BI}$ and
$C_{\mathcal{F}}$ stand for the homogeneous background
part and the ``matter'' correction encoding
inhomogeneities, respectively). Specifically, the
background part equals $C_{\rm BI} = \int_{\Sigma}
\rd^3 x |\det(E)|^{-1/2} \epsilon^{abc} E^i_a E^j_b
F_{ij\,c}$ where $F_{ij\,c}$ is the curvature (field
strength) of the connection $A^{i}_{a}$.

The quantization methods parallel those of LQG
\cite{rov}. In a first step, the constraint is
ignored. The basic objects promoted to operators are
the integrals of $A^i_a$ along straight lines
(holonomies) and those of $E_i^a$ along square
surfaces (fluxes). The resulting kinematical Hilbert
space is a product $\Hilk =
\bigotimes_{i=1}^3L^2(\bar{\re}_B,\rd\mu_B) \otimes
\Hilk_{\mathcal{F}}$ where $\bar{\re}_B$ is a Bohr
compactification of the real line and
$\Hilk_{\mathcal{F}}$ is a Hilbert space for the
inhomogeneous degrees of freedom.

In the next step, the constraint $C$ is promoted to
(and solved as) a quantum operator. For this, $C_{\rm
BI}$ has to be expressed in terms of holonomies and
fluxes. In particular, the term $F_{ij\,c}$ entering
$C_{\rm BI}$ is approximated by holonomies along small
rectangular loops. Since the limit of the loop
shrinking to zero does not exist in LQC, the
rectangular loop is fixed (following \cite{chiou}) by
the requirement that its physical area equals the
lowest nonzero eigenvalue of an area operator defined
in LQG.

In the matter part of $C$, the degrees of freedom
corresponding to gravitational wave modes
(conveniently scaled by a time-dependent function) are
represented via fields and momenta operators combined
into creation and annihilation operators
$(\hat{a}^\dagger_m,\hat{a}_m)$, $m=\pm 1,\pm 2,
\ldots$, with standard commutation relations
\cite{gauge}. Therefore they form the standard Fock
space $\Hilk_{\mathcal{F}}=\mathcal{F}$.

The final quantum constraint $\hat{C}$ is a difference
operator in all three coefficients $p_i$. In
principle, one can find the (generalized) states
annihilated by it, thus identifying the physical
Hilbert space $\Hilp$. This is indeed done in
\cite{gmm}. However, the representation of any state
of physical interest is complicated to the extreme by
the presence of an infinite number of degrees of
freedom. To be able to extract interesting physics out
of the system we appeal here to the so-called
classical effective dynamics.

To derive this effective description (see \cite{eff}
for details), one replaces the basic operators
(holonomies and fluxes) in the Hamiltonian constraint
by their expectation values, thus building back a
classical Hamiltonian. Nonetheless, some aspects of
the quantum theory are preserved by leaving the
lengths of holonomies as determined by the minimal
area requirement discussed above. The final form of
the constraint (after proper densitization) reads
\begin{eqnarray}
{\cal C}_G &=& -\frac{2}{\gamma^2}(\Theta_1 \Theta_2+
\Theta_1 \Theta_3+\Theta_2
\Theta_3)\nonumber\\
&+&\frac{G}{\gamma^2} (\Theta_2 + \Theta_3)^2
\frac{1}{|p_1|} H_{\rm int}^\xi + 32 \pi^2 G |p_1|
H_o^\xi,\label{eq:H-eff} \\ \Theta_i &=&
\frac{p_i\sqrt{|p_i|}}{M}
\sin(\frac{Mc^i}{\sqrt{|p_i|}}), \nonumber\\
H_o^\xi &=& \sum_{m=1}^{\infty} m \left[a_m^*a_m+
a_{-m}^*a_{-m}\right] ,\nonumber\\ H_{\rm int}^\xi &=&
\sum_{m=1}^{\infty}\frac{1}{m}
[a_m^*a_m+a_{-m}^*a_{-m}+a_{m}^*a_{-m}^* +
a_{m}a_{-m}].\nonumber
\end{eqnarray}
Here, $i=1$ denotes the direction in which there
exists spatial dependence, $G$ is Newton's constant
and $M=\sqrt{2\sqrt{3}\pi\gamma}\lPl$ ($\gamma$ is the
Immirzi parameter and $\lPl=\sqrt{G \hbar}$ the Planck
length). We note that the variables $\Theta_{2}$ and
$\Theta_{3}$ are constants of motion, a fact which
will considerably simplify the discussion of the
effective dynamics.

Given this effective constraint, it is straightforward
to derive the complete set of equations of motion,
namely the Hamilton-Jacobi equations. By integration,
one then obtains the time evolution of the system from
any initial point in the phase space.

At this stage, it is worth pointing out that the above
procedure does not take into account state-dependent
parameters (like dispersions and higher-order
quantities) which might significantly affect the
dynamics. However, comparison of this effective
approximation against the full quantum dynamics has
been carried out in simpler models (i.e., homogeneous
cosmologies with both massless and massive scalar
fields), showing that the corrections arising from
such parameters are negligible for the states of
physical interest (semiclassical at late times). This
fact has been confirmed analytically in the simplest
cosmological models \cite{taveras}. Although the
validity of the adopted effective description has not
been tested in the model considered here, the
commented results strongly support its ability to
reproduce the behavior of the quantum system quite
accurately.

We probed the dynamics of the effective system using
Monte-Carlo methods. First, a large population (c.a.
$10^4$ points) of initial points was selected randomly
in the phase space. In particular, apart from the
restriction to satisfy the $S^1$-translation
constraint, the values of $a_m$ were generated
randomly with Gaussian probability distribution
(centered at the origin). Owing to technical
limitations, simulations were restricted to a finite
number of nonvanishing modes (specifically, the cases
considered were $m\leq m_{\rm max} = 5,10,20$). To
find the dynamical trajectories, the initial value
problem consisting of the full set of equations of
motion plus the chosen initial data was integrated via
built-in adaptive methods of \emph{Mathematica}. The
analysis of the evolution was focused on two issues:
$(a)$ the existence of the bounce, and $(b)$ the
changes in the structure of the inhomogeneities
through the bounce.
\begin{figure}
\psfrag{Logpd}{$\log_{10}(p_2/l_{\rm Pl}^2)$}
\psfrag{Logpth}{$\log_{10}(p_1/l_{\rm Pl}^2)$}
\includegraphics[width=3.2in]{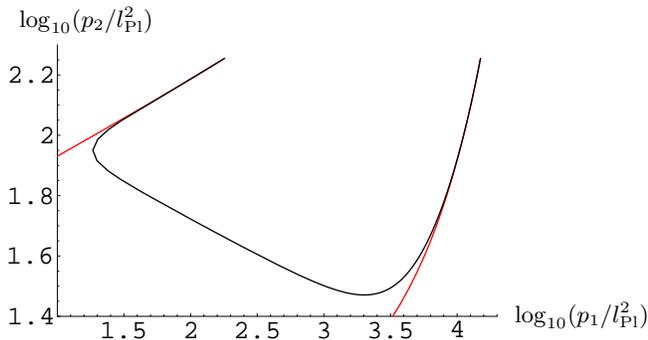}
\caption{A dynamical trajectory for the case $m_{\rm
max}=5$ (black) compared against the classical
trajectories to which it converges in the distant
future and past (red). Here, the value of $H_o^\xi$ at
the initial point ($p_1=p_2=180 l_{\rm Pl}^2$) is
$0.657 \hbar$. The Immirzi parameter is set to its
standard value, $\gamma=0.2375\ldots$}
\label{fig:traj}
\end{figure}

In order to address the first of these issues, the
evolution of the inhomogeneous universe was compared
against the evolution of its homogeneous counterpart,
i.e. the universe determined by the same initial data
for the homogeneous degrees of freedom and all
inhomogeneities set to vanish. The dynamical
trajectories obtained numerically confirm the presence
of the bounce in all three spatial directions (see
Fig. \ref{fig:traj}). Thus, the qualitative evolution
picture stays the same as in homogeneous models.
Besides, analytical studies show that the values of
$p_{2}$ and $p_{3}$ when they bounce do not depend on
the $a_m$'s; hence the bounces in the homogeneous
directions occur at the same values of $p_i$ as for
homogeneous universes. Finally, to check the behavior
of the bounce in the inhomogeneous direction, more
extensive analytical/numerical studies were performed.
Exploiting the properties of the equations of motion,
the investigation was carried out separately (and
using different methods) in two domains:
$\Theta_{2}\Theta_{3}<0$ and $\Theta_{2}\Theta_{3}>0$.

In the first case, it follows straightforwardly from
\eqref{eq:H-eff} that the presence of inhomogeneities
can only push the bounce away, that is, the bounce
happens at a larger value of $p_1$. Indeed, recalling
that $\dot{p}_{1} = \{p_{1},\cal{C}\}$ and taking into
account that $\dot{p}_{1}=0$ at the bounce, one gets
the constraint satisfied at that point,
\begin{equation}\label{eq:bounce} |p_{1}|^3 =
M^2 \frac{[F(p_{1},\Theta_{2},\Theta_{3}
;H_o^\xi,H_{\rm int}^\xi)-\Theta_{2}\Theta_{3}]^2}
{(\Theta_{2} + \Theta_{3})^2} ,
\end{equation}
where $F$ is a positive definite function such that
$F(p_1,\Theta_{2},\Theta_{3};0,0)=0$. The form of
\eqref{eq:bounce} implies immediately that, when
$\Theta_{2}\Theta_{3}<0$, the contribution of $F$
increases the value of $p_{1}$ (provided that
$\Theta_{2}$ and $\Theta_{3}$ remain the same). As a
consequence, the bounce in the inhomogeneous case
always happens at larger universe sizes than in the
homogeneous scenario.
\begin{figure}
\psfrag{H0}{$H_o^\xi / \hbar$} \psfrag{Logp}{$p_1 /
l_{\rm Pl}^2$}
\includegraphics[width=3.2in]{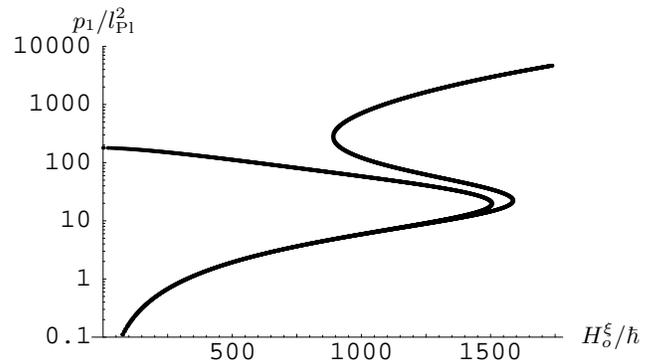}
\caption{The set of points satisfying relation
\eqref{eq:bounce} for the case $\Theta_2=3750 l_{\rm
Pl}^2$, $\Theta_3=2500 l_{\rm Pl}^2$, and $H_{\rm
int}^\xi/H_o^\xi=2\cdot 10^{-4}$. They form reflective
boundaries for the dynamical trajectories.}
\label{fig:bounce}
\end{figure}

The case $\Theta_{2}\Theta_{3}>0$ required a detailed
numerical analysis, since now the two terms in
brackets in the numerator of \eqref{eq:bounce} have
opposite signs. The behavior of the points satisfying
\eqref{eq:bounce} with respect to the magnitude of the
inhomogeneities is shown in Fig. \ref{fig:bounce}. A
feature which is worth noting is the existence of the
throat which allows, in principle, that the dynamical
trajectory may go ``down it'' through an infinite
sequence of bounces and recollapses, reaching the
singularity. However this would be a critical
trajectory (the set of initial data leading to such
evolution has zero measure in the phase space), thus
generically the universe will bounce at finite
$p_{1}$. Besides, except perhaps for a very small
subset of initial conditions (near the critical case),
the bounce happens at a value of $p_1$ above certain
bound, which is of similar order to the value found in
the absence of inhomogeneities, $p_{1}^o$. The bound
provided by our analysis is $\approx 0.13 p_{1}^o$.

Our study of the inhomogeneities was focused on
discussing how their energy distribution changes
through the bounce. This information is encoded in
$|a_m|(t)$. Since, for large $p_{1}$, these quantities
converge to well defined limits, it is particularly
interesting to investigate the corresponding amount of
asymptotic change for large universes, $\Delta|a_m| =
\lim_{t\to+\infty}|a_m|-\lim_{t\to-\infty}|a_m|$.
Actually, once the full set of initial data is
specified, the evolution of the universe (including
the inhomogeneities) is deterministic, so that the
value of $\Delta|a_m|$ is fixed. Nonetheless,
$\Delta|a_m|$ acquires a stochastic nature if one
restricts its attention to energies instead of
amplitudes for each of the gravitational wave modes,
thus ignoring the initial phases of the $a_m$'s.
Within the space of possible trajectories, on the
other hand, one can distinguish two differentiated
regimes: $(i)$ \emph{inhomogeneity dominated,} for
which the dynamics around the bounce is dominated by
the content of gravitational waves, and $(ii)$
\emph{near-vacuum}, for which those waves introduce
only small corrections to the vacuum Bianchi I
dynamics around the bounce. In case $(i)$ our
numerical analysis of a large population of universes
shows that, generically, $\Delta|a_m|$ is an
antisymmetric function of the initial phase of $a_m$.
Owing to this antisymmetry, the expectation value of
$\Delta|a_m|$ (namely, its average over the dependence
on initial phases) is ensured to vanish. In case
$(ii)$, a similar analysis shows that the
antisymmetric behavior of $\Delta|a_m|$ is generically
lost and its average becomes strictly positive.
Therefore, in the near-vacuum case the quantum
geometry effects around the bounce \emph{pump energy}
into the inhomogeneities.

It is worth noticing that, although our numerical
analysis has been performed in all cases for a finite
number of nonvanishing modes $m_{\rm max}$ (which
plays the role of an UV cutoff on the gravitational
waves) the results presented here do not change when
$m_{\rm max}$ increases. As a consequence, they remain
valid when the full Fock space of inhomogeneities is
considered.

The results explained above constitute the first
systematic analysis of the dynamics of inhomogeneous
spacetimes in LQC. In particular, they provide further
support to the bounce picture found earlier in
homogeneous scenarios. The study reported here, and
the methodology developed for it, paves the way for
future analyses of the effective physics derived from
LQG/LQC in more general spacetimes, opening an avenue,
e.g., for the discussion of the effects of quantum
geometry in the process of gravitational collapse and
black hole formation.

Let us conclude clarifying that the presented analysis
should not be treated as final. Firstly, our
inhomogeneous system has been investigated by means of
an effective theory which does not take into account
many quantum effects. To confirm the reliability of
the results, one ought to repeat the analysis in the
full quantum setting. Secondly, the method of
construction of the minimal area loop used to define
the quantum constraint $\hat{C}$ is not unique. There
exist several prescriptions for the construction,
giving different quantitative predictions of the
dynamics. Therefore, for robustness, an investigation
of the system constructed with those other
prescriptions is also desirable. Finally, one has to
remember that, in the studied system, there is only
one inhomogeneous direction; therefore the results
regarding change of inhomogeneities through the bounce
cannot be directly applied to the kind of models
considered in observational cosmology. To obtain
reliable results verifiable against observations, one
has to extend the analysis to models which have the
energy level degeneracy characteristic of a spherical
harmonics decomposition. This is the task that we plan
to accomplish in future work. \\

\noindent {{\bf Acknowledgments:} The authors are
grateful to J. Cortez, L.J. Garay, J.M.
Mart\'in-Garc\'ia, and especially M. Mart\'in-Benito.
This work was supported by the Spanish MICINN Project
FIS2008-06078-C03-03 and the Consolider-Ingenio 2010
Program CPAN (CSD2007-00042). D.B. acknowledges
financial aid by the FPI Program of Madrid Regional
Government, and T.P. by the I3P Program of CSIC
(together with the ESF) and the Foundation for Polish
Science grant {\it Master}.

\end{document}